\documentclass[11pt,leqno,fleqn]{article}
\title{Comments on "The Casimir effect upon a single plate"}
\author{A. Kwang-Hua Chu} %\thanks{This paper has not been
\date{Department of Physics, Xinjiang University,
Urumqi 830046, PR China}
\begin{document}           % End of preamble and beginning of text.
%%\renewcommand{\baselinestretch}{1}
%\baselinestretch=1        %\intextskip=6mm
\maketitle                 % Produces the title.
\begin{abstract}
We make remarks on Hoodbhoy's paper [{\it J. Phys. A: Math. Gen.} 38 (2005) 10253-10256]
by pointing out the single plate considered could be elastic and there will be deformations
once the net (one-sided) force being upon the plate which will change the position-dependent potential imposed by Hoodbhoy. \newline

\noindent
PACS number: 11.10.-z
\end{abstract}
%----------------------------------------------------------------------
\doublerulesep=6mm        %\parskip=12 pt
\baselineskip=6mm
\oddsidemargin+1mm         %%\evensidemargin-12mm
\bibliographystyle{plain}
Hoodbhoy just showed that even a single conducting plate may experience
a non-zero force due to vacuum fluctuations. The force thus estimated is positive, increases with the distance of the plate away from the origin and is non-analytic in the strength of the external potential. In [1] the origins of this force lie in the
change induced by the external potential in the density of available quantum
states (in fact, the difference in the density of normal modes above and below the plate, induced by the position-dependent external potential, is the responsible mechanism [1]). As Hoodbhoy commented that this force has the same origin as the Casimir
effect (externally imposed boundary conditions on a freely fluctuating electromagnetic field lead to the famous Casimir force between conducting surfaces separated by some small distance [2]), i.e. is a manifestation of the zero-point fluctuations of a quantum field.
Similar effect of vacuum fluctuations upon a single boundary could also be traced in [3]
(please refer to [4] for other interesting cases or developments).
Therein the Casimir force on a single surface immersed in an inhomogeneous medium
was reported. To be specific, if $\Delta(x - a)$ represents a semi-penetrable
thin plate placed at $x = a$. In the limits :
$\Delta(x - a) \rightarrow \delta(x - a)$; the scalar field obeys a Dirichlet
boundary condition, $\phi = 0$, at $x = a$.  In all the cases Jaffe and Scardicchio
have studied [3] they found that the Casimir force on the plate points in the
direction opposite to the
force on the quanta of $\phi$ : it pushes the plate toward higher potential,
hence their use of the term {\it buoyancy}. \newline
Hoodbhoy considered a real scalar field $\phi(x)$ described by the Lagrangian
$L = (\partial_{\mu} \phi)^2 /2 -V (x)\,\phi^2 /2$, where $V (x)$ is an externally
prescribed field.
By way of mocking up a constant force directed towards a fixed centre
at $x = 0$, Hoodbhoy chose $V (x) = b|x|$ with $b > 0$ and $-\infty < x < \infty$.
Intuitively speaking, as a scalar photon rises it loses energy and undergoes a
redshift. The Dirichlet condition $\phi(a) = 0$ will be said to represent a single
{\it conducting plate} placed above the origin at a height $a$. For a
translationally invariant potential, the forces on both sides of the plate would
cancel. But, with a position-dependent potential, this would not be true.
Hoodbhoy obtained the force as the space-space component of the canonical
energy-momentum tensor $T^{\mu\nu}$, or after the integration (cutting off the
integral at the lower end with a value $K^2 \sim |a|b$ in which case
$T^{xx} \sim |a|b$)
: $T^{xx}= \hbar c \eta^{2/3} f(\eta)/a^2$. $y = \kappa^2 +(x/a)\eta^{1/3}$ where
$\eta = b\,a^3$ and $k = i\,b^{1/3}\kappa$. $k$ comes from the Fourier transform [1].
  \newline
Based on the present author's experiences, some remarks could be made below (cf Figure 1).
If the plate is rigid enough, then it can sustain the force. Otherwise,
once the plate is elastic, the net (one-sided) force upon the plate will induce an elastic stress or strain per unit area which is related to a deformation to the plate.
Thus the plate cannot support a constant force or the position-dependent external potential
will be varied depending on the deformation at the instant the force is effective upon the plate. Meanwhile, the plate will be moving if there is no minimum-energy states.
The possible stable state is a bubble-like plate : after the imposing of a net force
upon the outer (spherical) plane the deformations thus obtained shrink the inner volume
or increase the inner pressure (which is isotropic and homogenized)
to balance the outer force field and will reach a local final force-balance in vacuum. \newline
One illustrative example for the realistic materia of the plate (not only elastic) :
most recently in [5] strict limits for the Casimir interaction were proven that
apply to all causal and linear materials, including both bulk and multilayer
structures. Henkel  and Joulain illustrated these results by computations of the
Casimir pressure, considering materials with frequency-dispersive response
functions like those encountered in effective medium theories [6].
Henkel and Joulain also derived power law exponents and prefactors and find that
a strongly modified Casimir interaction is possible in a range of distances
around the resonance wavelength of the response functions.
{\it Acknowledgements.} The author is partially supported by the Starting Funds of
2005-XJU-Scholars. \newline

\vspace*{20mm}

\vspace{3mm}
\setlength{\unitlength}{1.00mm}   %\psdraft
\begin{picture}(120,54)(0,-16)
\thicklines

\put(10,0){\line(1,0){100}}
\put(10,-4){\vector(0,1){4}}
\put(60,-4){\vector(0,1){4}}
\put(110,-4){\vector(0,1){4}}
\put(35,-4){\vector(0,1){4}}
\put(22.5,-4){\vector(0,1){4}}
\put(85,-4){\vector(0,1){4}}
\put(97.5,-4){\vector(0,1){4}}
\put(47.5,-4){\vector(0,1){4}}
\put(72.5,-4){\vector(0,1){4}}

\put(60,35){\circle{15}}
\put(60,24){\vector(0,1){4}}
\put(60,46){\vector(0,-1){4}}
\put(49,35){\vector(1,0){4}}
\put(71,35){\vector(-1,0){4}}
\put(68,43){\vector(-1,-1){3}}
\put(52,43){\vector(1,-1){3}}
\put(52,27){\vector(1,1){3}}
\put(68,27){\vector(-1,1){3}}

\put(2,-15){\makebox(0,0)[bl]{\small Fig. 1 \hspace*{1mm} Schematic
diagram of a single plate and floating bubble-like plate.}}
\put(2,-19){\makebox(0,0)[bl]{\small The later (spherical shape)
is much more stable then the former (flat plate).}}
%\put(10,-2){\makebox(0,0)[bl]{\small \hspace*{12mm} %prescribed as {\it
%%vertices} or
\end{picture}

\end{document}